\documentclass[sigconf]{acmart}
\renewcommand\footnotetextcopyrightpermission[1]{} 
\usepackage{times}  
\usepackage{helvet}  
\usepackage{courier}  
\usepackage{natbib}  
\usepackage{caption} 
\usepackage{algorithm}
\usepackage{algorithmic}
\usepackage{comment}
\usepackage{booktabs}
\usepackage{multirow}
\usepackage{caption}
\usepackage{subcaption}
\usepackage{subcaption,siunitx,booktabs}
\usepackage{xhfill}
\AtBeginDocument{%
  \providecommand\BibTeX{{%
    \normalfont B\kern-0.5em{\scshape i\kern-0.25em b}\kern-0.8em\TeX}}}

\begin{document}

\newtheorem{altehypothesis}{Alternative Hypothesis}
\newtheorem{nullhypothesis}{Null Hypothesis}

\newcounter{opti}
\renewcommand{\theopti}{Optimization}


\title{Price-guided user attention in \\ large-scale E-commerce group recommendation}

\author{Yang Shi}
\affiliation{%
  \institution{Rakuten Group, Inc.}
  \streetaddress{800 Concar Dr}
  \city{San Mateo}
    \state{CA}
  \country{USA}}
\email{yang.shi@rakuten.com}

\author{Young-joo Chung}
\affiliation{%
  \institution{Rakuten Group, Inc.}
  \streetaddress{800 Concar Dr}
  \city{San Mateo}
    \state{CA}
  \country{USA}}
\email{youngjoo.chung@rakuten.com}

\renewcommand{\shortauthors}{Shi and Chung}

\begin{abstract}
 Existing group recommender systems utilize attention mechanisms to identify critical users who influence group decisions the most. We analyzed user attention scores from a widely-used group recommendation model on a real-world E-commerce dataset and found that item price and user interaction history significantly influence the selection of critical users. When item prices are low, users with extensive interaction histories are more influential in group decision-making. Conversely, their influence diminishes with higher item prices. Based on these observations, we propose a novel group recommendation approach that incorporates item price as a guiding factor for user aggregation. Our model employs an adaptive sigmoid function to adjust output logits based on item prices, enhancing the accuracy of user aggregation. Our model can be plugged into any attention-based group recommender system if the price information is available.  We evaluate our model's performance on a public benchmark and a real-world dataset. We compare it with other state-of-the-art group recommendation methods. Our results demonstrate that our price-guided user attention approach outperforms the state-of-the-art methods in terms of hit ratio and mean square error.\end{abstract}

\ccsdesc[500]{Information systems~Recommender systems}

\keywords{Recommender systems, Group recommender systems, Attention mechanism}


\maketitle

\section{Introduction}
On E-commerce websites, many promotional campaigns and advertisements target user segments. User segments are created based on their attributes: demographics, location, interests, and shopping behavior. Marketers of E-commerce platforms utilize user segments to create tailored ad contents that can resonate with the interests and preferences of each segment. For example, an E-commerce website conducts a special campaign for outdoor gear for user groups like young adults and families with children. For the young adults' segment, the advertiser may recommend ski equipment, and for families with children, they may recommend large-sized tents. 

Recently, many approaches have been proposed to recommend items to groups of users~\cite{Masthoff2015,O’Connor2001,Hu_Cao_Xu_Cao_Gu_Cao_2014}. Previous works mainly studied two categories of groups: persistent groups and ephemeral groups. Persistent groups consist of  fixed, social-connected members (e.g., families watching movies)~\cite{groupim}. Ephemeral groups consist of users who are together briefly (e.g., attending a conference)~\cite{10.1145/2959100.2959137}. However, E-commerce user segments belong to neither of the two groups. They are not ephemeral because they does not change frequently. They are also different from fixed groups because users in user segments do not influence and interact with others in the group. Another main difference is that the size of the user segment groups in E-commerce is significantly bigger than the group size in datasets for academic research. Existing methods with complex architectures, such as multi-view graphs~\cite{10.1145/3459637.3482081,10.1145/2792838.2800190}, may achieve good performance, but may be computationally inefficient for E-commerce group recommendation.


To understand how the existing group recommender model performs on the E-commerce user segments, we ran the widely-used group recommender model (AGREE~\cite{agree}). We analyzed how it aggregates users to recommend items to groups. We investigated the users who influenced the recommendation results most and found that item price plays a pivotal role in deciding such users. We utilized item and user content information and built a simple yet novel attention-based model: Price-Guided User Attention (PGUsA). We designed a novel attention module that takes in the item price and user activity history statistics and outputs a value determining how much this user's opinion matters to the group's decision. In summary, our contributions are:
\begin{itemize}
    \item We analyzed critical factors in determining users' importance in E-commerce group recommendations.
    \item We proposed the novel group member aggregation model: PGUsA, which utilizes users' and items' content information. 
    \item We performed extensive experiments on a public benchmark and a real-world dataset to show the effectiveness and efficiency of our model.    
\end{itemize}


\begin{figure}[!t]
  \centering
  \includegraphics[width=0.7\linewidth]{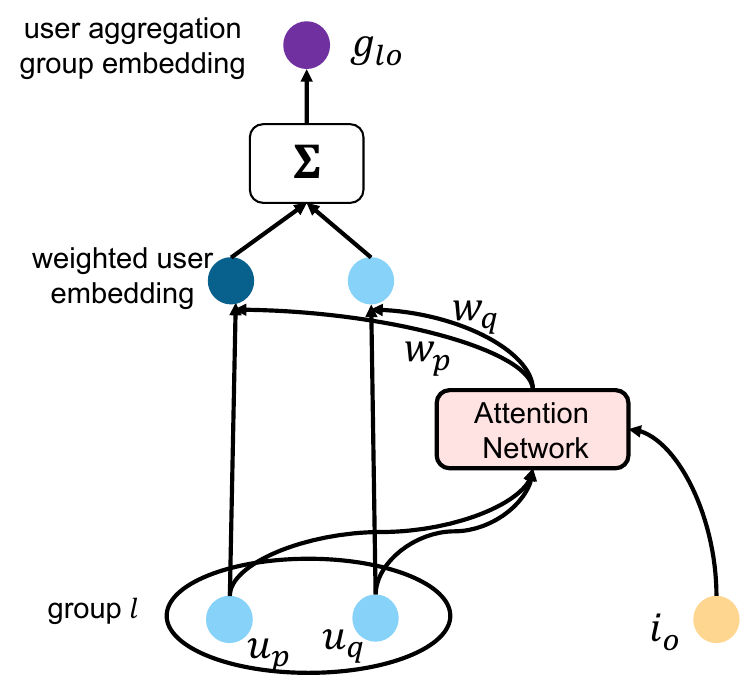}
  \captionof{figure}{User preference aggregation in AGREE. $u_p$ and $u_q$ are user representations in group $l$, $i_o$ is item $o$'s representation, $g_{lo}$ is group $l$'s representation for item $o$, $w_p$ and $w_q$ are the user aggregation weights.}
  \label{fig:agree}
\end{figure}

\section{Related work}
Existing group recommendation methods can be separated into two categories. One is to learn group preferences directly, and the other is to learn preferences aggregated from users. With enough training observations, we can learn group preference directly through collaborative filtering (CF) or other traditional recommender systems (RS) methods. However, in a real-world setting, group activities are much fewer than individual activities. Learning directly from group interactions will be sub-optimal. Thus, more works focused on user preference aggregation. These models can also be divided into two types of approaches: 

\textit{Two-step score aggregation} These approaches have two steps: first, the model learns individual preferences towards items and then aggregates all scores of individuals who belong to the same group to compute groups' preferences~\cite{grr,linas}. Usually, the aggregation methods are hand-crafted: the average~\cite{10.1007/978-3-319-03844-5_65} and the least misery~\cite{10.14778/1687627.1687713}, et cetera.
These models are easy to implement. However, the hand-crafted aggregation rules may not fully reflect the group's consensus.

\textit{End-to-end score aggregation.}  
Neural models became a popular method for learning group profiles. Group embeddings are aggregations of user embeddings with learnable weights ~\cite{10.1145/3331184.3331251, 8807225, com,ul}. For example, AGREE~\cite{agree} is proposed to learn to select influential users in the group using a vanilla attention mechanism. Similar works are OGA~\cite{9101842} with self-attention, and GroupIM~\cite{groupim}, which explores different aggregation methods and maximizes the mutual information between group and user. To capture the  more profound roles of each member in group decision-making, pre-trained language models~\cite{gbert}, hypergraphs~\cite{hiegraph,cuberec,9679118,Zhang2021DoubleScaleSH}, and multi-view learning~\cite{10.1145/3397271.3401064} are proposed for group recommendation. The state-of-the-art(SOTA) model ConsRec~\cite{consrec} includes member-level,
 item-level tastes, and group-level inherent preferences aggregations with hyper-graphs to explore consensus behind group behavior.
\begin{figure}[!t]
    \centering
    \includegraphics[width = 0.85\linewidth]{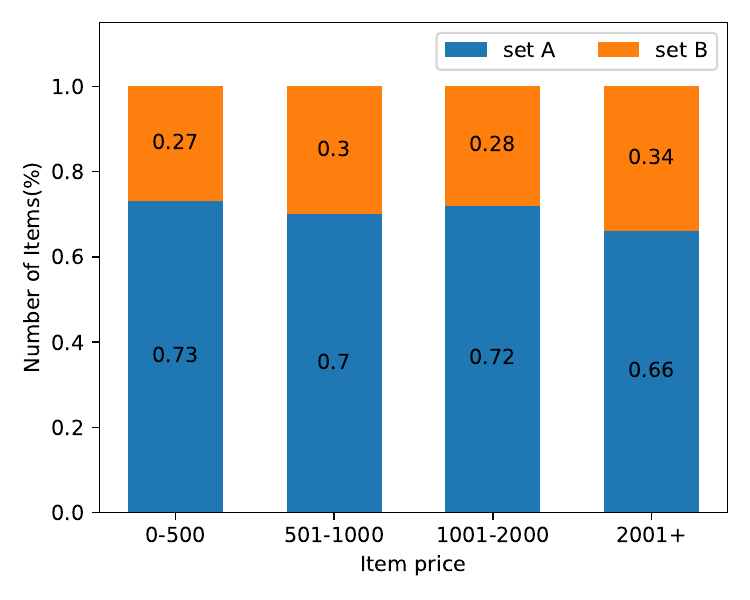}
    \caption{Item distribution segmented by item price. Set A is a set of items where the most influential user is a frequent buyer. Set B contains the rest of the items.  }
    \label{fig:pie}
\end{figure}

\section{Price-Guided User Attention }
\subsection{Hypothesis: Frequent buyers are more influential for cheap items}
\label{sec:hp}

Attention-based group RS models give users different attention weights (influence) to create the group embedding for each item. The influence level is learned during the training. 
To identify the most influential users, we trained AGREE model on real-world e-commerce data and obtained the users' attention scores on the test data. For each purchased test item, we selected 19 non-purchased items, and the model ranked these 20 items.  We then investigated users' attention weights (i.e., $w_p$ and $w_q$ in Figure~\ref{fig:agree}) only when the correct item was ranked as the top candidate. 
To understand the correlations between the user's attention weight and purchase frequency, we first defined a frequent buyer as someone who has made more than twice the average number of purchases per user within their group for simplicity. Then we divided the corrected-ranked items into two sets. In set A, the most influential user (i.e., the user with the highest attention weight) for the item is a frequent buyer. Otherwise, the item is placed in set B. Figure~\ref{fig:pie} shows the item distribution in sets A and B. From the figure, on the one hand, we see that the most influential buyers are typically frequent buyers, as indicated by the higher ratio of items in Set A across all price ranges. On the other hand, as the item price increases, the ratio of items in Set A tends to decrease, indicating that frequent buyers are less likely to be the influential users in the group for expensive items.

The observation makes us wonder if \textbf{\textit{Frequent buyers are more influential for cheap items?}} To verify our assumption, we did chi-square tests on the following hypotheses:
\begin{nullhypothesis} There is no association between buyer frequency (frequent vs. non-frequent) and being the most influential user for items priced below the 10th percentile.
\end{nullhypothesis}

\begin{nullhypothesis} There is no association between buyer frequency (frequent vs. non-frequent) and being the most influential user for items priced above the 90th percentile. 

\end{nullhypothesis}
Here, we choose 10 percentile ($P_{10}$) price  and 90 percentile ($P_{90}$) price as examples of cheap and expensive item prices. The two categories we considered in the chi-square tests are C1: influential users are frequent buyers (purchase more than two times of the average number of  purchases per user in the group), and C2: influential users are non-frequent buyers. In the Null hypothesis, these two groups should have an equal chance to be the most influential user in the group. Table~\ref{tab:hp1} shows the calculated chi-square value of 79.51, which lies in the reject region. Thus, we conclude that when the item price is low, the group does not give an equal likelihood for frequent/non-frequent buyers to be the most influential user. On the other hand, in Table~\ref{tab:hp2}, the chi-square value is less than the critical value, so we cannot reject null hypothesis 2, which assumes an equal probability for frequent/non-frequent buyers to be influential.

\begin{table}[!t]
    \centering       
   \caption{Hypothesis 1 chi-square test with critical value 3.84 (use 5\% significant level)}
\begin{tabular}{c|c|c|c|c}
    \toprule
        & C1  & C2  & $x^2_c$ & results\\
        \midrule
     Expected  & 489.5 & 489.5 & \multirow{2}{*}{  79.51} & \multirow{2}{*}{ Reject Null}\\
     Observed &629& 350 & & \\
     \bottomrule
    \end{tabular}
   
\label{tab:hp1}
\end{table}

\begin{table}[!t]
    \centering
   \caption{Hypothesis 2 chi-square test with critical value 3.84 (use 5\% significant level)}
\begin{tabular}{c|c|c|c|c}
        \toprule
         & C1  & C2  &   $x^2_c$ & results\\
         \midrule
     Expected  & 16.5 & 16.5& \multirow{2}{*}{0.76 } & Not enough evidence  \\
     Observed & 14 & 19 &&to reject Null  \\
     \bottomrule
    \end{tabular}
 
          \label{tab:hp2}
\end{table}

\subsection{Notations}
The group recommendation task is defined as follows: suppose we have $n$ users,
$s$ groups
, and $m$ items 
. Each group contains a set of users, and the group size is the number of users in the group. We also  have past user-item interactions and group-item interactions. The interactions can be explicit or implicit feedback. We use $X = [x_{ab}]_{n \times m}$ and $Y = [y_{ab}]_{s \times m}$ to represent the interactions, respectively. The final goal is: given a group, find items that this group may be interested in.

\subsection{Price-Guided User Attention (PGUsA)}
\begin{figure}[!t]
\centering
    \includegraphics[width = 0.8\linewidth]{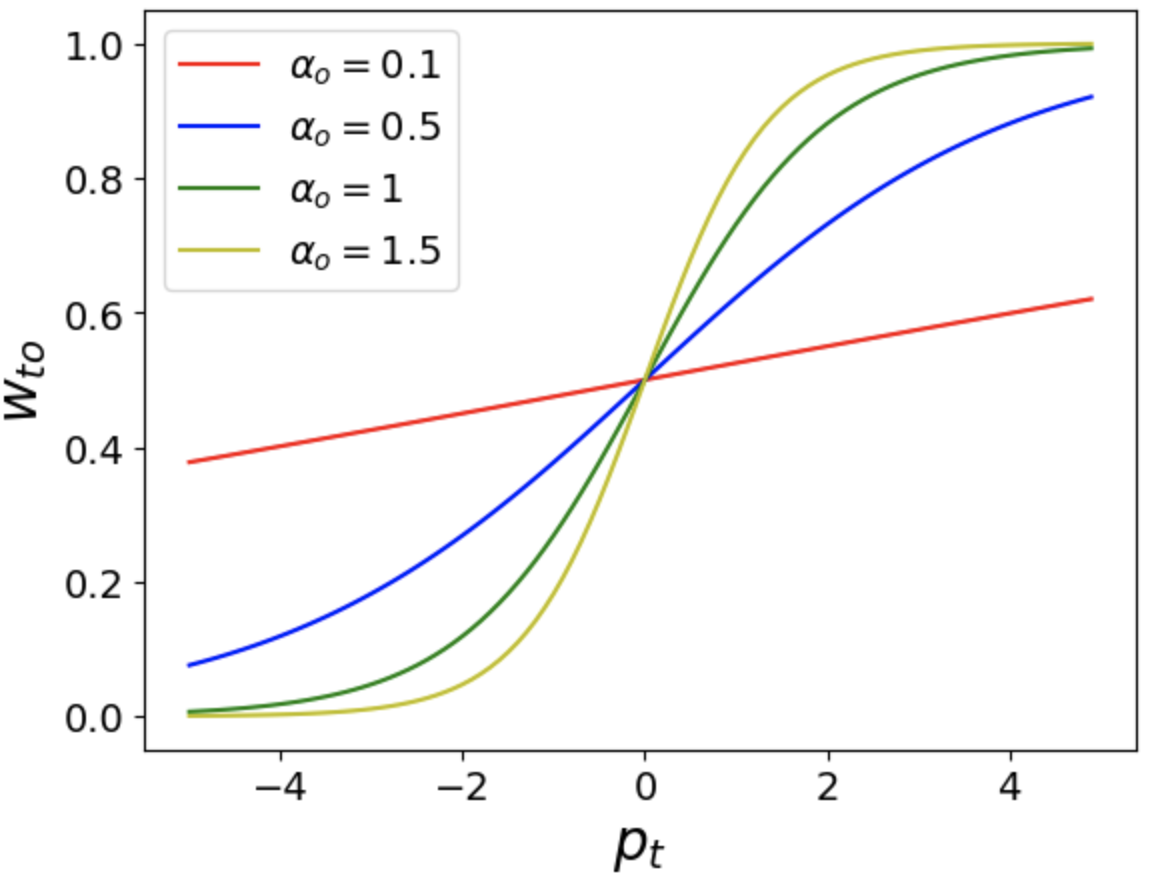}
    \captionof{figure}{Example of adaptive sigmoid functions. The output weight changes based on item price-related parameter $\alpha_o$ and user purchase frequency-related parameter $p_t$.}
    \label{fig:adaptsig}
\end{figure}

\begin{figure}[t]

  \centering
  \includegraphics[width=0.7\linewidth]{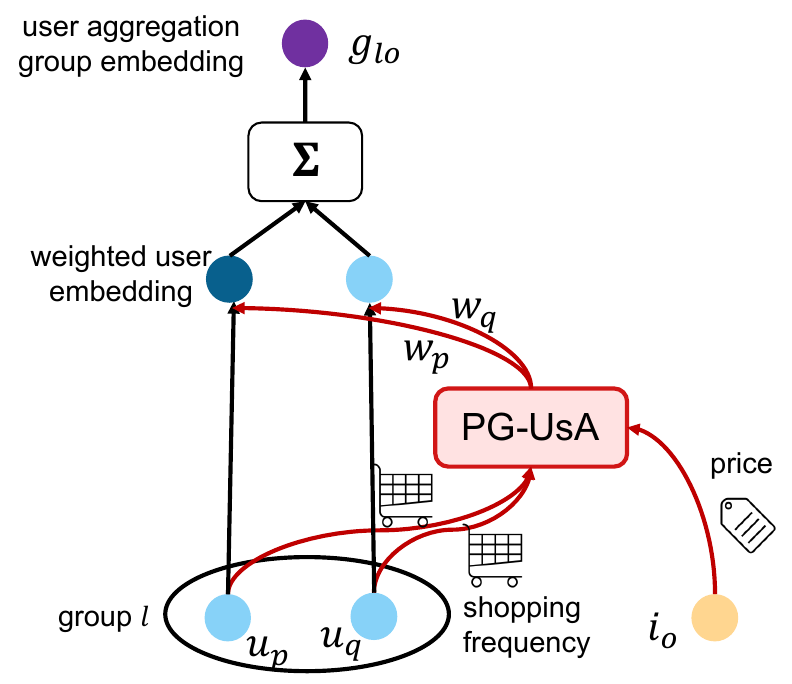}
  \captionof{figure}{Price-guided user attention (PGUsA) module. We utilize user shopping frequency and item price information to adjust the user aggregation weight.}
  \label{fig:pgusa}
\end{figure}

Given the observations from Section~\ref{sec:hp}, we introduce a Price-Guided User Attention module for group recommendation. The main challenge is to reflect item price and user interaction frequency information into the user weights. We adopt an adaptive sigmoid function to  solve the problem:
\begin{equation}
\label{eqn:attention}
    w_{to} =  \frac{\beta}{1+e^{- \alpha_o p_t}}
\end{equation}
where $w_{to}$ is user $t$'s weight towards item $o$, $\alpha_o$ is inverse proportion to the item $o$'s price, $p_t$ is proportion to user $t$'s purchase frequency, and $\beta$ is a scalar. In this model, on the one hand, when the item price is low, $\alpha_o$ is high, which makes the adaptive sigmoid function has a steep slope, more frequent buyers will have drastically more significant attention than less frequent buyers. On the other hand, when the item price is high, $\alpha_o$ is low, the sigmoid function has a smaller slope, and the weights differences between frequent buyers and non-frequent buyers are smaller. An illustration of the adaptive sigmoid function is shown in Figure~\ref{fig:adaptsig}. The PGUsA module is shown in Figure~\ref{fig:pgusa}. With the PGUsA module, we create a new group embedding $g_{lo}$ as follows:
\begin{equation}
    g_{lo} = \sum_{u_t \in \text{group } l} w_{to} u_t
\end{equation}
Note that this embedding is item dependent.
With the help of PGUsA, we can have better group representations and, therefore, better group recommendation performance.

\subsection{Model}
\begin{figure}[t]
  \centering
  \includegraphics[width=.7\linewidth]{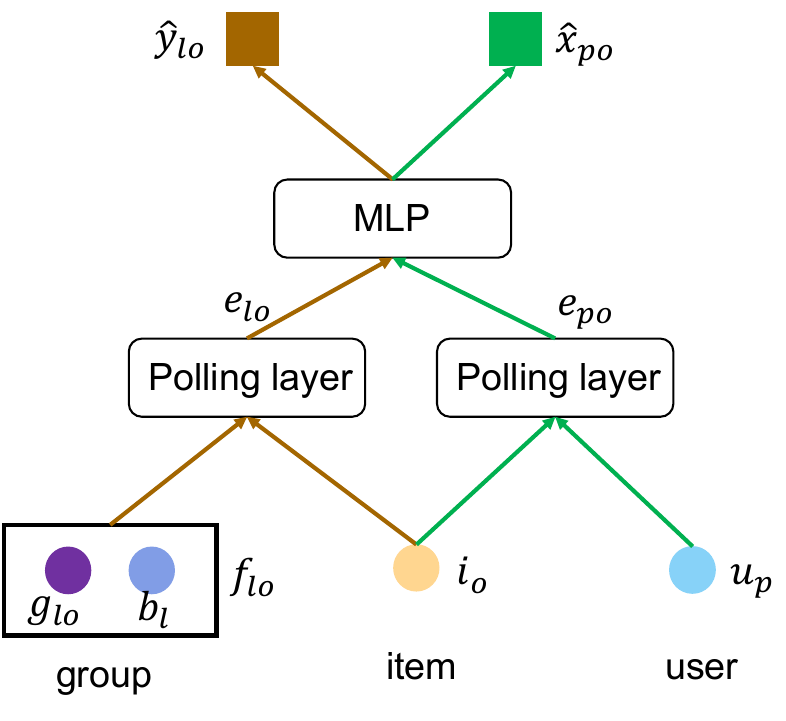}
  \captionof{figure}{AGREE model with PGUsA group embedding $g_{lo}$ (shown in purple). }
  \label{fig:model}
\end{figure}

PGUsA can be added to any user aggregation-based group recommendation model if user/item side information is available. Here, we add PGUsA to AGREE and show the model in Figure~\ref{fig:model}. For simplicity, in the following sections, AGREE with PGUsA model is referred to as the PGUsA model. The model's inputs are group and item embeddings, and the output is a predicted score for the group-item pair $\hat{y}_{lo}$. 

\textbf{Group embedding:} Following the AGREE, the group embedding $f_{lo} = g_{lo}+b_l$. It has two components: user aggregation group embedding $g_{lo}$ and independent group embedding $b_l$. The first embedding $g_{lo}$ is item dependent, whereas the second is not. 

\textbf{Pooling layer:} Similar to the AGREE, this pooling layer concatenates three embeddings: group embedding, item embedding, and an element-wise product of the group and item embeddings. The output of the pooling layer is:
\begin{equation}
 e_{lo} =\text{concatenate}(f_{lo}, i_o, f_{lo} \odot i_o )  
\end{equation}
The pooling vector is then fed-forward  to a multi-layer perception (MLP). The output of the MLP is the predicted score $\hat{y}_{lo} = \text{MLP}(e_{lo})$.

The model's parameters are also shared to predict user-item scores $x_{po}$ since the additional user-item interaction information may help learn better group representations. Instead of using group embeddings in the pooling layer, we use user embeddings when predicting user-item scores.

\subsection{Optimization} 
\label{sec:opt}

We choose different objective functions based on whether the interaction is implicit or explicit.
We adopt regression-based pairwise loss~\cite{agree}:  $\mathcal{L} = \sum_{l,o,u}(\hat{y}_{lo}-\hat{y}_{lu}-1)$ for implicit feedback predictions.
In this case, we select a negative item $u$ that the group does not interact with for each positive group-item sample. 
The goal is to separate the positive and negative sample scores. For explicit feedback like review ratings, we adopt mean square error (MSE) loss: $\mathcal{L} = \sum_{l,o}(y_{lo}-\hat{y}_{lo})^2$.
Detailed optimization settings can be found in Section~\ref{sec:expmodel}.

\section{Experiments}

In this section, we present experiments to answer the following research questions (RQs):
\begin{itemize}
    \item RQ1: How does our proposed PGUsA perform compared with other group recommendation methods?
    \item RQ2: How can PGUsA improve the SOTA model?
    \item RQ3: How does PGUsA affect gross merchandise value (GMV) improvement?
\item RQ4: How efficient is our method compared with other group recommendation models?
\item RQ5: How does hyperparameter $\beta$ affect the model performance?
\end{itemize}

\subsection{Datasets and Experimental Setup}
\subsubsection{Datasets}
We use two datasets for the experiments: Epinion\footnote{http://www.epinions.com} review dataset and  Ichiba\footnote{https://www.rakuten.co.jp} purchase dataset. Epinion.com~\cite{epinion} was a general consumer review site that allowed users to rate items and also provided the option for users to ''trust'' each other. The Ichiba dataset is a real-world E-commerce dataset that contains user purchase histories from Category A on the Ichiba website, which are implicit feedback.  These datasets were selected because they include item price information, which is crucial for our analysis.

\textit{Group definition}: In the Epinion dataset, groups are formed based on social trust: each group contains people who trust at least one of the others in the group. In the Ichiba dataset, groups are formed by age:  upto 18,19-22,23-29,30-34,35-39,40-44,45-49,50+. Average group sizes are 24 and 338,247 for the two datasets, respectively.

For the Ichiba dataset, we created age-based groups because demographic information is commonly used to create user segments. However, for the Epinion dataset, we didn’t have user attribute information. Therefore, we formed groups of users who trust each other based on a trust network. The assumption is that these user segments would be provided by clients (e.g., target users in a social media group). 

\textit{Group-item interaction definition}: In Epinion, the group's preference for an item is the average ratings of all users. In Ichiba, due to the sparsity of the dataset, only less than 15\% of the items were purchased twice in the same group. We assume the group interacted with the item if at least two users in the group purchased the item. To further evaluate the model with different data densities, we create a denser Ichiba dataset, Ichiba-s. In this dataset, we define the group-item interaction as positive if at least 101 users in the group purchased the item. Because of the constraint, group 18- is filtered out. The average group size is 2,057, which is more than 100 times smaller than the  Ichiba dataset's group size.

\textit{Item price and user review/purchase frequency information:}
We normalize the price inversely between 0.01 and 1 (0.01 means most expensive) and normalize the purchase frequency between 0 and 5 (5 means most frequently). 

\begin{table*}[!t]
\caption{Dataset statistics}
    \label{tab:data}
    \centering
        \scalebox{1}{
  \begin{tabular}{ccccccc}
    \toprule
    Dataset& \#User & \#Item & \#Group & \#U-I interaction (train) & \#G-I interaction (train) & G-I interactions (test)\\
    \midrule
Epinion &  13,623& 23,816& 584 &37,970 & 21,894& 7,342\\
Ichiba &  2,733,458& 535,262 &8 &5,667,546& 452,593 & 7,140 \\
Ichiba-s &  14,397& 158 &7 & 8,112 & 146  & 40 \\
  \bottomrule
    \end{tabular}
    }

\end{table*}

\begin{table*}[!t]
\centering
  \caption{Different models' performances on Epinion, Ichiba, and Ichiba-s (* means the PGUsA's improvements are significant)}
  \begin{tabular}{c||c|c|c|c|c|c|c|c}
    \toprule
   \multirow{2}{*}{Models} & \multicolumn{2}{c|}{Epinion}& \multicolumn{3}{c|}{Ichiba} & \multicolumn{3}{c}{Ichiba-s} \\
   \cline{2-9}
    &MSE& MAPE &{HR@1} & {HR@10} &  {NDCG@10}&{HR@1} & {HR@10} &  {NDCG@10}  \\
    \midrule
    
Popular & NA  & NA&0.0008 & 0.0073 & 0.0035 & 0.0000 & 0.0004&0.0002 \\
AGREE & 1.4298  &  0.3688   &0.7408 & 0.9574 & 0.8560 & 0.0250 & 0.4750 & 0.1841
\\
NCF & 1.3951  &   0.3633  &0.7389& 0.9588 & 0.8555 & 0.0500 & 0.4500 & 0.1998\\
\hline
  NCF-AVG & 1.5737&  0.3730    &0.7933 &0.8092& 0.8024 & 0.2500 & 0.3750&0.3013\\
    NCF-EXP&1.5275  &   0.3817  &\textbf{0.7950}& 0.8113& 0.8043  &  \textbf{0.2750} & 0.3500&0.3108\\
    PGUsA  &  $\textbf{1.2788}^{*} $   & $\textbf{0.3613}^{*} $ &0.7417& $\textbf{0.9620}^{*}$ & \textbf{0.8578} & 0.1250 & $\textbf{0.6750}^{*}$& $\textbf{0.3308}^{*}$
\\
   \bottomrule
\end{tabular}

    \label{tab:ichiba}
\end{table*}

\begin{table*}[!t]

    \centering
            \caption{Consrec model performance on Epinion, Ichiba, and Ichiba-s (* means the ConsRec+PGUsA improvements are significant)}
    \begin{tabular}{c||c|c|c|c|c|c|c|c}
    \toprule
   \multirow{2}{*}{Models} &  \multicolumn{2}{c|}{Epinion}& \multicolumn{3}{c|}{Ichiba} & \multicolumn{3}{c}{Ichiba-s} \\
   \cline{2-9}
    &MSE &MAPE&{HR@1} & {HR@10} &  {NDCG@10}&{HR@1} & {HR@10} &  {NDCG@10}     \\
    \midrule
       ConsRec &  8.9733 &   0.6571    &0.9471 & 0.9668 & 0.9541 & 0.3250 &  0.8750 & 0.5563\\
       ConsRec+PGUsA & \textbf{8.9303}&    \textbf{0.6397 }     & $\textbf{0.9598}^{*}$ & \textbf{0.9790} & \textbf{0.9667}  &$ \textbf{0.9750} ^{*}$&$\textbf{0.9750}^{*}$& $\textbf{0.9750}^{*}$\\
          \bottomrule
    \end{tabular}

            \label{tab:consrecichiba}
\end{table*}

\subsubsection{Models}

We compare PGUsA's recommendation performance with the following models:
\begin{itemize}
    \item \textbf{Popularity}~\cite{pop}: This is a non-personalized model. The popularity of an item is its number of interactions in the training set. We recommend the most popular items to all groups in the test.
    \item \textbf{AGREE}~\cite{agree}: AGREE utilizes a vanilla attention mechanism to aggregate user preferences and builds an NCF-like model to predict group and user preferences. 
    \item \textbf{NCF}~\cite{NCF}: In this neural network-based CF method, we treat groups as virtual/additional users and train them together with users.
    \item \textbf{NCF-AVG} and \textbf{NCF-EXP}: In these models, we only train with user-item interactions and then integrate users' preferences to obtain the group's preferences.  AVG averages everyone's score. EXP is a weighted average, and the weight is based on the user's expertise:  users with higher purchase frequency are more expertise than users with low purchase frequency. 
    \item \textbf{ConsRec}~\cite{consrec}: ConsRec is the state-of-the-art group recommendation model. It includes  member-level aggregation, item-level tastes, and group-level inherent preferences with hyper-graphs.
\end{itemize}

\subsubsection{Implementation details} 
\label{sec:expmodel}

With Ichiba datasets, since the interactions are binary, PGUsA, AGREE, and ConsRec use regression-based pairwise loss (from Section \ref{sec:opt}). Followed by the original papers~\cite{NCF}, NCF-based models use binary cross-entropy loss. The positive:negative sample ratio for training is 1:1.  With Epinion data, all models use MSE loss. We use embedding size 8, batch size = 256 in all experiments. The hyperparameters $\beta =5 $ in PGUsA are selected from grid-search. All embedding layers use uniform Xavier initialization; other linear layers use normal distribution initialization. The optimizer is RMSprop with learning rate $1e^{-4}$ and $1e^{-3}$ for Ichiba and Epinion experiments, respectively. All models are implemented using Pytorch 1.12.1, and all experiments are done on Nvidia V100.
\subsubsection{Evaluation metrics}
For the Epinion dataset, we randomly divided the groups into an 80:20 train:test split. For the Ichiba dataset, we trained the model using a dataset from 8/1/2020 to 8/23/2021, and tested it on purchase made between 8/24-2021 and 8/31/2021. 
Following previous settings \cite{agree, consrec}, we adopt Hit Ratio (HR) and Normalized Discounted
Cumulative Gain (NDCG) as evaluation metrics for Ichiba datasets. As a standard evaluation protocol to reduce evaluation computation time~\cite{he2017neural}, for each test item that the group interacted with, we selected 19 non-purchased items.  We let the model rank these 20 items instead of rank over all items from the dataset. For Epinion dataset, we use MSE and Mean Absolute Percentage Error (MAPE). Notice that all experiments are done in the offline setting.
We train models three times (with different initialization) and mark results with asterisks if they are significantly better than the second-best ones using a t-test with a significance level of 0.1.
\subsection{Overall Performance (RQ1)}
We compare PGUsA's performance with other baselines in Table~\ref{tab:ichiba}. We have the following observations:
\begin{itemize}
    \item PGUsA performs best among all models on Epinion, Ichiba, and Ichiba-s datasets most of the time. It has more than 10\% relative MSE improvement compared to AGREE on the Epinion dataset. Even though NCF-EXP outperforms PGUsA at HR@1 on Ichiba datasets, NCF-EXP was more than 15\%  and 80\% worse than PGUsA at HR@10 on Ichiba and Ichiba-s datasets, respectively. Overall, the results verified the effectiveness of our PGUsA model.
    \item User-based NCF integration models' performances are not stable. It is difficult to justify their performances. Nevertheless, between AVG and EXP aggregations, EXP performs better. EXP also used user purchase history as input, implying the importance of the user purchase history feature.
    \item Popular model's results are the worst among all baselines. However, in other papers~\cite{agree,consrec}, Popular model performs relatively well. The difference is because of the dataset size differences. Compared to other public group recommendation datasets such as Mafengwo\footnote{http://www.mafengwo.cn} and CAMRa2011\footnote{http://2011.camrachallenge.com/2011}, Epinion and Ichiba datasets are much bigger and more sparse. Only recommending popular items will hurt the recommendation performance.
    \item Even though Ichiba-s has a more strict group-item interaction definition, requiring at least 101 users in the group to interact with the item compared to other datasets, our model still outperforms other models on Ichiba-s. Between Ichiba-s and Ichiba, because of the lack of training samples, HR on Ichiba-s is not as good as the HR on Ichiba.
\end{itemize}

\begin{figure}[!t]
    \centering
    \includegraphics[width=\linewidth]{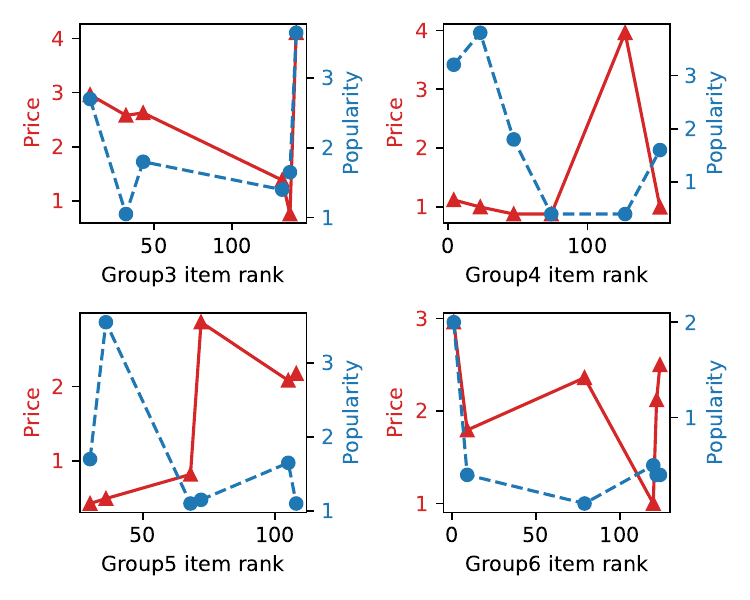}
    \caption{Truly interacted items' prices and popularities versus the recommendation ranks using NCF-EXP model on Ichiba-s. Values are normalized for privacy concerns.}
    \label{fig:pp-ncf}
\end{figure}

\subsection{Enhancement to SOTA ConsRec (RQ2)}
ConsRec, the state-of-the-art group recommendation model, combines different views of aggregations: member-level hypergraph aggregation $G^m$, group-item level bipartite graph aggregation $G^i$, and group-group level weighted graph aggregation $G^g$. ConsRec outperformed almost any previous group recommendation models. We add the PGUsA module to ConsRec to see if we can improve the performance. Specifically, we sum the PGUsA-weighted group embedding $g_{lo}$ with the three-view aggregation: $g^{new}_{lo} =  G_{lo}^g+G_{lo}^i + G_{lo}^m +g_{lo}$. The results are in Table~\ref{tab:consrecichiba}. PGUsA boosted the result by more than 1\% at HR@1 and HR@10 on Ichiba dataset. Surprisingly, ConsRec failed to perform well on the Epinion dataset, compared to other base models. We believe it is because of the design of ConsRec's prediction layer.  ConsRec computes dot products between group and item embeddings if the task requires explicit feedback to avoid a dead ReLU problem. This design may not be compatible with MSE loss. Nevertheless, Consrec+PGUsA improved the performance.

\begin{figure}[!t]
    \centering
    \includegraphics[width=\linewidth]{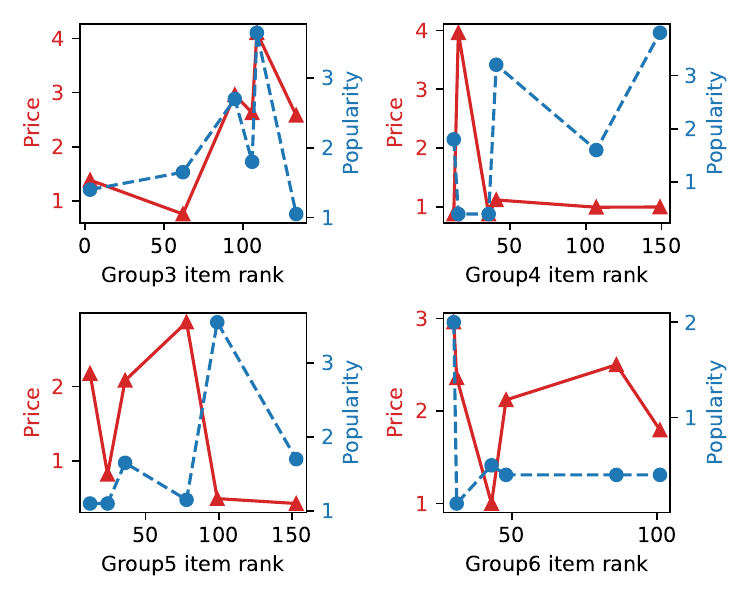}
    \caption{Truly interacted items' prices and popularities versus the recommendation rank using PGUsA model on Ichiba-s.}
    \label{fig:pp-ours}
\end{figure}

\begin{figure}[!t]
    \centering
    \includegraphics[width=0.95\linewidth]{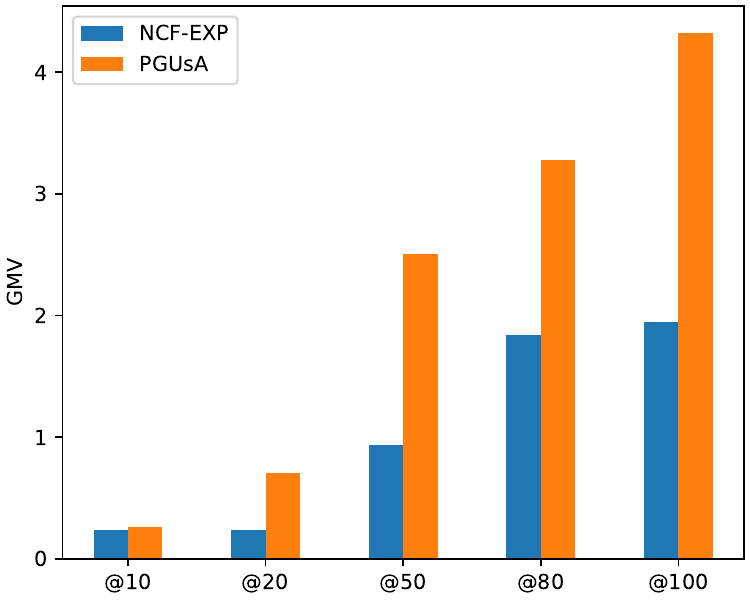}
    \caption{GMV increase over recommendation ranks. Values are normalized.}
    \label{fig:gms}
\end{figure}

\begin{figure*}[!t]
  \centering
  \includegraphics[width=0.85\linewidth]{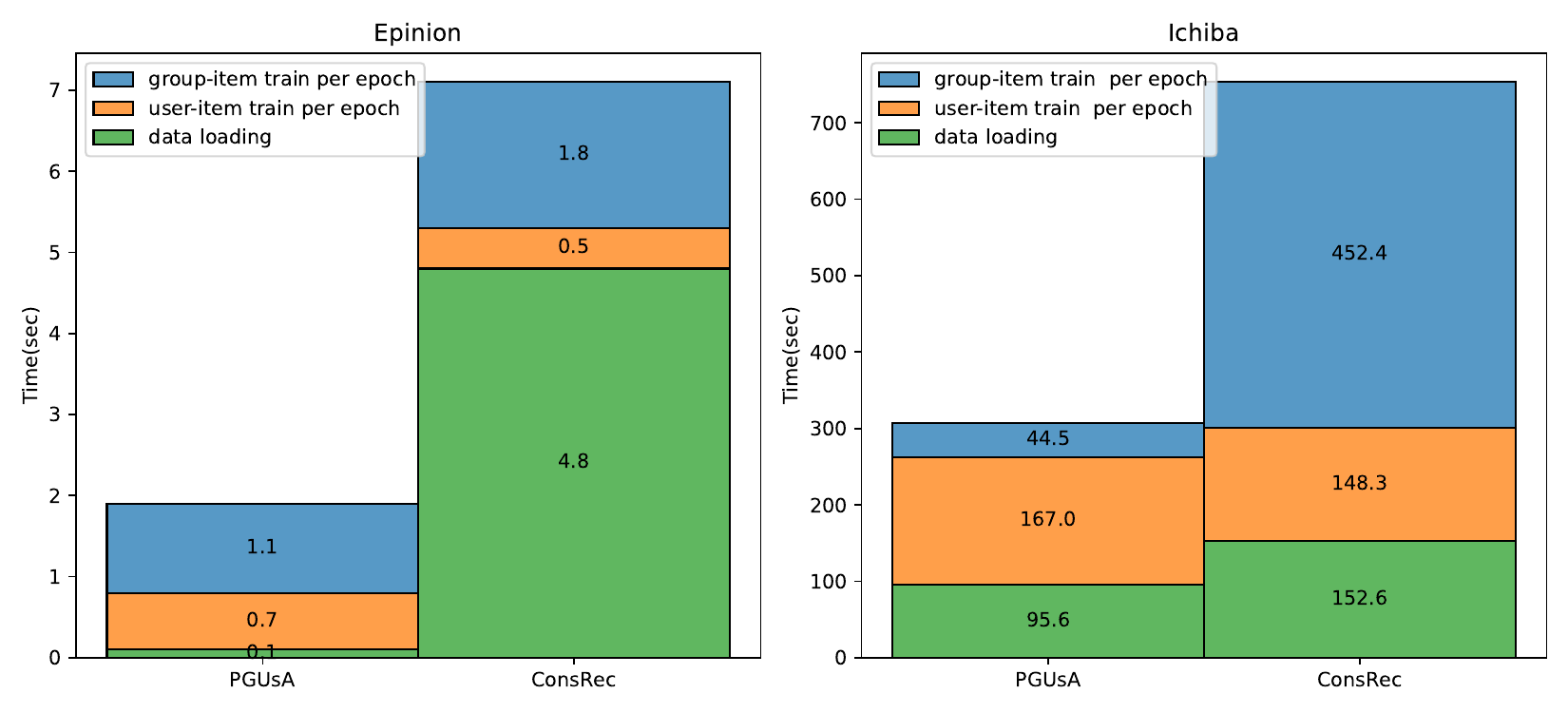}
  \captionof{figure}{Training time analysis on Epinion and Ichiba dataset.}
  \label{fig:time_epinion}
\end{figure*}

\begin{table*}[!t]
\centering
 \caption{$\beta$'s effect on group recommendation performance}
  \centering
  \begin{tabular}{c||c|c|c|c|c|c|c|c}
    \toprule
   \multirow{2}{*}{$\beta$} & \multicolumn{2}{c|}{Epinion}& \multicolumn{3}{c|}{Ichiba} & \multicolumn{3}{c}{Ichiba-s} \\
   \cline{2-9}
    &MSE &MAPE& {HR@1} & {HR@10} &  {NDCG@10}&{HR@1} & {HR@10} &  {NDCG@10}  \\
    \midrule
    1 &2.9509&0.3943 &0.7195& 0.9531&0.8416 &0.0250 & 0.5000& 0.1958\\
    5 & \textbf{1.2788} & \textbf{0.3613}&\textbf{0.7417}& \textbf{0.9620} & \textbf{0.8578} & \textbf{0.1250} &\textbf{0.6750}& \textbf{0.3308}
\\
10 &2.4490 & 0.3905&0.7001 &0.9434 & 0.8272& 0.0750& 0.6000&0.2650\\
   \bottomrule
\end{tabular}
 
    \label{tab:beta}
\end{table*}

\subsection{GMV improvement (RQ3)}
In addition to HR and NDCG, GMV is essential in evaluating model performance in real-world E-commerce platforms. GMV = $\sum_{i} \text{price}_i * \text{purchase}_i$, where $\text{price}_i$ is the $ith$ item's price and $\text{purchase}_i$ is the $ith$ item's number of sales. To further understand how the model's recommendation affects the GMV improvement, we rank all items for each group. We then examine their predicted ranks, prices, and popularities (i.e., the number of buyers) within each group. Note that the same item would have different rankings for different groups. To reduce the computation cost, we use Ichiba-s dataset. We check PGUsA and NCF-EXP because these two models are the top-2 models in terms of HR and NDCG from Table~\ref{tab:ichiba}. We show truly interacted items' prices and popularities in the sequence of predicted rank from high to low in Figure~\ref{fig:pp-ncf}, ~\ref{fig:pp-ours}. Due to the page limit, we show group 3 to group 6's results. 
The observations are:
\begin{itemize}
    \item NCF-EXP model is more likely to recommend popular items first. In Figure~\ref{fig:pp-ncf}, the popularity line (blue dots) decreases as each group's recommendation rank decreases. It can be an issue if the recommender system only recommends popular items to the groups.
    \item PGUsA model recommendation does not follow trivial patterns, as shown in Figure~\ref{fig:pp-ours}. For groups with younger buyers, such as group 3, the model is more likely to recommend cheap items first. For older groups, such as group 5 and 6, the model will recommend more expensive items first.
\end{itemize}
Different recommendation rankings will lead to different GMV growth. Figure~\ref{fig:gms} shows the GMV increase over predicted item recommendation ranks. The GMV growth using NCF-EXP is smaller than PGUsA's. For instance, at rank 50,  GMV obtained from PGUsA is more than twice the GMV from NCF-EXP. The conclusion is that PGUsA not only improves HirRatio or NDCG but also increases GMV. 

\subsection{Efficiency of our model (RQ4)}
We evaluate model efficiency based on model size and training time. 
We calculate the number of parameters needed for each model: On the Ichiba dataset, PGUsA, AGREE, and ConsRec each have around 26M parameters, while NCF-based models have around 52M parameters. NCF~\cite{NCF} models require more parameters  because they are a fusion of Generalized Matrix Factorization (GMF) and MLP. GMF and MLP must learn separate embeddings for each user/item, doubling the model size. Even though ConsRec, the SoTA model, performs better on the Ichiba and Ichiba-s datasets, its complex model results in longer training times when compared to PGUsA. Figures~\ref{fig:time_epinion} shows Consrec takes $2.8$x total training time each epoch than PGUsA on Ichiba data. ConsRec's data loading time is longer than ours because it has to initialize all the graphs. Additionally, group-item training is more time-consuming because training the hyper-graphs takes longer than the MLP.

\subsection{Hyperparameter analysis (RQ5)}
The hyperparameter $\beta$ in Equation~\ref{eqn:attention} influences the sensitivity of the attention score. Higher $\beta$ leads to a more significant difference in the attention score of low/high frequent buyers. We show the performance w.r.t. different $\beta$ in Table~\ref{tab:beta}. We observe that better results can be obtained when $\beta$ is 5. 

\section{Conclusions}
This paper proposes a price-guided user attention model (PGUsA) for group recommendations in E-commerce. Our analysis shows that item price is a critical factor in identifying influential users in group recommendations. Frequent buyers are more influential for low-priced items, while their influence diminishes for high-priced items. We designed an adaptive sigmoid module to learn the relationships between item price, user behavior, and user weight in the group. Experiments demonstrate that PGUsA outperforms baseline models like AGREE and can enhance the performance of state-of-the-art models.

\newpage
\bibliographystyle{ACM-Reference-Format}
\bibliography{sample-base}


\end{document}